\documentclass[pra,twocolumn,aps,amssymb,footinbib]{revtex4-1}
  \usepackage[dvips]{color}

\usepackage{amsfonts}
\usepackage{amsmath}
\usepackage{amssymb}
\usepackage{graphicx}%
\providecommand{\U}[1]{\protect\rule{.1in}{.1in}}

\begin{document}

\preprint{APS/123-QED}

\title{Topologically Protected Quantum State Transfer in a Chiral Spin Liquid}
\author{N. Y. Yao$^{1\dagger*}$, C. R. Laumann$^{1,2\dagger}$, A. V. Gorshkov$^{3}$, H. Weimer$^{1,2}$, L. Jiang$^{3}$,  J. I. Cirac$^{4}$, P. Zoller$^{5}$, M. D. Lukin$^{1}$}

\affiliation{$^{1}$Physics Department, Harvard University, Cambridge, MA 02138}
\affiliation{$^{2}$ITAMP, Harvard-Smithsonian Center for Astrophysics, Cambridge, MA 02138}
\affiliation{$^{3}$Institute for Quantum Information, California Institute of Technology, Pasadena, CA 91125}
\affiliation{$^{4}$Max-Planck-Institut fur Quantenoptik, Hans-Kopfermann-Strase 1, Garching, D-85748, Germany}
\affiliation{$^{5}$Institute for Quantum Optics and Quantum Information of the Austrian Academy of Sciences, A-6020 Innsbruck, Austria}
\affiliation{$^{\dagger}$These authors contributed equally to this work}
\affiliation{$^{*}$e-mail: nyao@fas.harvard.edu}

\date{\today}
\begin{abstract}

Topology plays a central role in ensuring the robustness of a wide variety of physical phenomena. Notable examples range from the robust current carrying edge states associated with the quantum Hall and the quantum spin Hall effects to proposals involving topologically protected quantum memory and quantum logic operations. Here, we propose and analyze a topologically protected channel for the transfer of quantum states between remote quantum nodes. In our approach, state transfer is mediated by the edge mode of a chiral spin liquid.  We demonstrate that the proposed method is intrinsically robust to realistic imperfections associated with disorder and decoherence. Possible experimental implementations and 
applications to the detection and characterization of spin liquid phases are discussed.
\end{abstract}

\maketitle
The decoherence of both quantum states and quantum channels represents a major hurdle in the quest for the realization of scalable quantum devices~\cite{Chuang95,ChuangBook}. 
Several avenues are currently being explored to address these important challenges. For example, quantum repeater protocols are expected to improve the fidelity of quantum state transfer -- the fundamental building block of quantum communication \cite{Briegel98,Kimble08}.  Similarly, quantum error correction can significantly extend the lifetime of quantum memories and suppress the errors associated with quantum logic operations \cite{Preskill98, Gottesman98}. The practical realization of these technologies, however, requires a high level of quantum control that is as yet, not experimentally accessible.  An alternative paradigm to achieving protected quantum states is provided by topology; indeed, if such states can be stored in the topological degrees of freedom of certain exotic states of matter, they become intrinsically robust against local noise \cite{Kitaev01, Dennis02, Kitaev03, Ioffe02, DasSarma06, Nayak08}. 

The implementation of robust long-lived quantum memories can also be achieved by encoding quantum bits in appropriately chosen physical degrees of freedom. For example, the natural isolation of nuclear spins immunizes them from the environment and makes them an exceptional candidate for the storage of quantum information \cite{Childress06, Dutt07, Balasubramanian09, Neumann10b, Morton08, Kurucz09}. Such solid-state spin qubits can be locally coupled with high fidelities, enabling the realization of few-bit quantum registers \cite{Dutt07, Neumann10b}. However, spatially remote registers interact extremely weakly; thus, in this context, the challenge of scalability is shifted to the development of quantum channels capable of connecting remote registers in a robust and noise-free fashion \cite{Taylor05,Bose07,YJG10}. 

This article describes a novel approach to the realization of intrinsically robust quantum channels and exploits topological protection to enable high-fidelity quantum information transport. We envision quantum state transfer between remote spin registers to be mediated by a 2D system composed of interacting spins. Specifically, the spin system is tuned into a gapped chiral spin liquid phase, which harbors a fermionic edge mode.  The prototype of this specific chiral spin liquid is the gapped B phase (CSLB) of the Kitaev honeycomb model \cite{Kitaev06}. Although such a phase is best known for its non-Abelian vortex excitations, here, by operating at finite temperatures below the gap, we make use of  its Majorana fermionic edge mode as a topologically protected quantum channel. Moreover, we discuss possible applications of our protocol for the spectroscopic characterization of spin liquid states \cite{Hermele09}. 

\section*{Approach to Topologically Protected State Transfer}

Our approach to quantum state transfer is schematically illustrated in Fig.~1. Quantum information is encoded in a two-qubit spin register, with each qubit capable of being individually manipulated.  The register is coupled to the edge of a two-dimensional spin droplet, whose elements we assume cannot be individually addressed but can be globally ``engineered'' to create a spin liquid state in the CSLB phase. The transfer protocol proceeds by mapping the quantum information stored in the left-hand spin-register onto the chiral edge mode of the droplet. The resulting wavepacket traverses the edge before retrieval at the remote register. 


A distinct feature of our protocol, as compared with previous approaches \cite{Bose07, YJG10, Christandl04, Clark05, Fitzsimons06}, is the fundamental robustness of the quantum channel. The chiral nature of the fermionic edge mode ensures that destructive backscattering during state transfer is highly suppressed; moreover, the characteristic (linear) dispersion of the edge-mode ensures that wave packet distortion is minimized. Finally, we demonstrate that our approach is remarkably insensitive to disorder and decoherence affecting both the bulk and edge of the droplet. Although any spin system with a stable CSLB-like phase can potentially mediate topologically protected state transfer (TPST), to illustrate the microscopic mechanism responsible for such state transfer, we turn initially to a particular model and will later generalize our analysis to include the effects of disorder, additional interactions, and decoherence.

\begin{figure}
\centering
\includegraphics[width=3.4in]{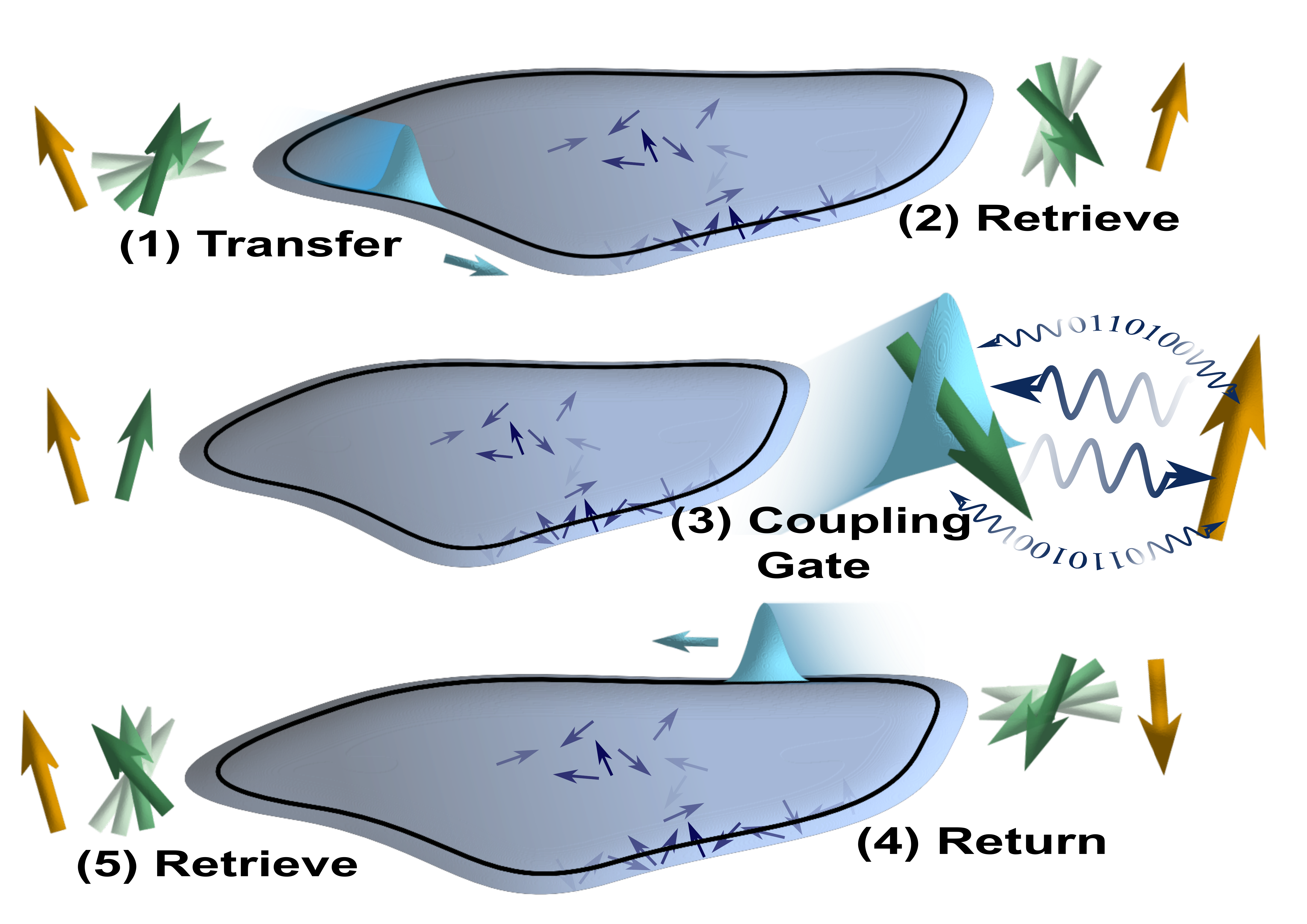}
\caption{\label{fig:Fig1} {\bf Schematic representation of topologically protected state transfer $|$} The grey droplet represents a 2D array of interacting spins tuned into the CSLB phase. Quantum spin-registers composed of a transfer qubit (green) and a memory qubit (gold) are arranged around the edge of the 2D droplet and coupling between them occurs through the chiral edge mode. (1) By mapping the quantum information onto a fermionic wave-packet (blue) traveling along the edge, the quantum state can be transferred to a remote register. The wavepacket travels only in the direction of the blue arrow; this chirality prevents mode localization and destructive backscattering. At a specified time at the remote register location, the coupling is turned on and the wavepacket is captured (2). Given an ancillary memory qubit and local register manipulations, a two-qubit gate (3) can be performed before the quantum state is transferred back to the original register and stored (4-5). This allows for universal computation between the memory qubits of spatially separated registers. }
\end{figure}

\section*{TPST on the Decorated Honeycomb}

We now consider a specific exactly solvable spin-$1/2$ model which supports robust TPST \cite{Yao07}. Within this Yao-Kivelson model, the spins are situated on a triangular-decorated honeycomb lattice as depicted in Fig.~2 \cite{Yao07}.  The associated Hamiltonian naturally generalizes the Kitaev model \cite{Kitaev03} and features a chiral spin liquid ground state (CSLB phase), 
\begin{equation}
H_0 = \frac{1}{2}\sum_{\genfrac{}{}{0pt}{}{x,x'}{links}} \kappa \sigma^{x}_{i} \sigma^{x}_{j} + \frac{1}{2}\sum_{\genfrac{}{}{0pt}{}{y,y'}{links}} \kappa \sigma^{y}_{i} \sigma^{y}_{j} + \frac{1}{2}\sum_{\genfrac{}{}{0pt}{}{z,z'}{links}} \kappa \sigma^{z}_{i} \sigma^{z}_{j},
\end{equation}
\noindent where $\vec{\sigma}$ are Pauli spin operators ($\hbar = 1$). The model may be solved  by introducing four Majorana operators, $\{ \gamma^{0}, \gamma^{1},\gamma^{2},\gamma^{3} \}$ for each spin, as shown schematically in Fig.~2a and by representing the spin algebra as: $\sigma^{x} = i \gamma^{1} \gamma^{0}$, $\sigma^{y} = i \gamma^{2} \gamma^{0}$, $\sigma^{z} = i \gamma^{3} \gamma^{0}$ \cite{Kitaev06,Yao07}. The Majorana operators are Hermitian and satisfy the standard anticommutation relation $\{ \gamma^{l},\gamma^{m} \} = 2\delta_{lm}$. The Hilbert space associated with the physical spin is a two-dimensional subspace of the extended four-dimensional Majorana Hilbert space; thus, we must impose the gauge projection, $P =  \frac{1+D}{2}$, where $D= \gamma^1 \gamma^2 \gamma^3 \gamma^0$ \cite{Kitaev06}.

Transforming to Majorana operators results in the extended Hamiltonian
\begin{equation}
H^{\gamma} = \frac{i}{4} \kappa \sum_{i,j} \hat{U}_{i,j} \gamma_{i}^{0} \gamma_{j}^{0},
\end{equation}
\noindent where $\hat{U}_{i,j} = i\gamma^{\alpha}_{i} \gamma^{\alpha}_{j}$ ($\alpha$ depends on the type of $ij$-link) for $ij$ connected and zero otherwise; these $\hat{U}_{i,j}$ correspond to the boxed Majorana pairs illustrated in Fig.~2a. Remarkably each $\hat{U}_{i,j}$ commutes with the Hamiltonian and with all other $\hat{U}_{l,m}$, implying that the extended Hilbert space can be divided into sectors corresponding to static choices of $\{ U_{i,j} = \pm 1\}$ \cite{Kitaev06, Yao07}.


The choice of $\{ U_{i,j} \}$ yields a Hamiltonian which is quadratic in the $\gamma^0$ Majorana operators; from the perspective of these Majoranas, $U_{i,j}$ is a static background $\mathbb{Z}_2$ gauge field.  The physical states are sensitive only to the flux of the gauge field, $w(p) = \prod_{ij \in \partial p} U_{i,j}$, where $p$ represents a plaquette, $\partial p$ is its boundary and $ij$ is oriented according to the arrows in Fig.~2b \cite{Yao07}. For any link with $U_{i,j} =+1$, this orientation can also be interpreted as the direction in which a $\gamma^0$ Majorana hops in order to accumulate a $\pi/2$ phase. The ground state flux sector of the model has $w(p) = +1$ for all plaquettes, corresponding to $\pi$ phase around the dodecagonal plaquettes and $\pi/2$ phase around the triangular plaquettes, as shown in Fig.~2b.  The $\pi/2$ phase around the triangular plaquettes indicates the breaking of time-reversal symmetry necessary for a chiral ground state.  Alternate flux sectors contain plaquettes with vortex excitations defined by $w(p) = -1$. In general, such vortices are energetically gapped by $\Delta_v$, but the energy and dynamics of vortices near the edge are controlled by the details of the boundary.  

In each flux sector, the associated Majorana Hamiltonian can be diagonalized through a unitary transformation $Q$ such that $\kappa \sum_{i,j} Q_{k,i} (i U_{i,j}) Q^{*}_{k',j} = \delta_{kk'}\epsilon_{k}$, yielding $H^{\gamma} = \frac{1}{2} \sum_{k=-N/2}^{N/2} \epsilon_{k} c_{k}^{\dagger} c_{k}$, where $c_{k} = \frac{1}{\sqrt{2}}\sum_{j} Q_{k,j} \gamma^{0}_{j}$, $N$ is the number of spins on the lattice, and the index $k$ is ordered according to energy.  Owing to particle-hole symmetry, $\epsilon_k = -\epsilon_{-k}$ and $c_{k}^{\dagger} = c_{-k}$; thus, by restricting to $k > 0$, 
\begin{equation}
H^{\gamma} = \sum_{k>0} \epsilon_{k} (c_{k}^{\dagger} c_{k}-\frac{1}{2}),
\end{equation}
where $c_k$ and $c_{k}^{\dagger}$ satisfy Dirac anticommutation relations. Diagonalization of the ground state flux sector on a cylinder yields three bulk fermion bands, energetically gapped by $\Delta_b$, as shown in Fig.~3~\cite{Yao07}. At the edge, the fermionic quasiparticles form gapless chiral modes which are guaranteed by the nontrivial Chern number of the bulk fermion bands.

\begin{figure}
\centering
\includegraphics[width=3.4in]{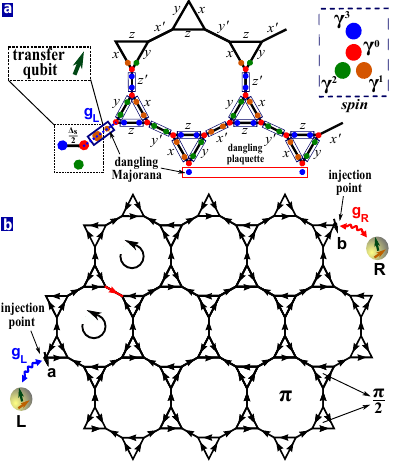}
\caption{\label{fig:Fig2} {\bf Coupling between spin-register and the droplet edge $|$}  (a) Schematic representation of the generalized Kitaev Hamiltonian on the decorated honeycomb lattice. Spins are represented by four Majorana operators; spin-spin interactions become products of the four Majoranas living on each link. Boxed spins correspond to the $\hat{U}_{i,j}$ operators which determine the effective hopping associated with the $\gamma_0$ Majoranas. Coupling (dashed line) between the quantum register and the 2D droplet can occur at any edge vertex with an unpaired Majorana; the Majorana flavor determines the form of the spin-spin interaction that introduces the desired additional hopping. (b) Schematic representation of the ground state flux configuration. Each arrow represents a Majorana hopping of $i$, yielding $\pi$ phase (oriented CCW) around the dodecagonal plaquettes and $\pi/2$ phase around the triangular plaquettes. Vortex excitations (circular arrows) correspond to the flipping of a $U_{ij}$ link (red arrow and link), which yields $w(p) = -1$ in the two adjacent plaquettes.  Quantum registers corresponding to a transfer qubit (green) and a memory qubit (gold) are shown coupled into the droplet (injection points) at two different dangling edge spins.}
\end{figure}

\subsection*{Spin-Register Coupling to a Chiral Edge}

We now consider the addition of spin qubits, which can be individually manipulated and read out, to the edge of the exactly solved model with open boundary conditions \cite{Dutt07, Neumann10b}. Each edge spin with coordination two has an uncoupled Majorana operator, which we term dangling as depicted in Fig.~2a. We can extend the definition of vortices to include the dangling plaquettes defined by the $U_{i,j}$ links between dangling Majoranas, as shown by the red rectangle in Fig.~2a.  These dangling vortices are completely
decoupled from the fermions and lead to a large degeneracy of the model. However, generic perturbations will lift this degeneracy by gapping out these dangling vortex states; in this situation, as we later describe, the control of dangling vortices at the injection point will become important. 



To illustrate TPST, we consider the full Hamiltonian $H_{T} = H_0 + H_{int}$ where $H_{int}$ characterizes the coupling between the two spin-registers (termed $L$ and $R$) and dangling spins at the edge of the droplet (Fig.~2b), 
\begin{equation}
H_{int} = -\frac{\Delta_S}{2} (\sigma^{z}_{L} + \sigma^{z}_{R}) + g_L \sigma^{\beta}_L \sigma^{\beta}_{a} + g_R \sigma^{\eta}_R \sigma^{\eta}_{b}.
\end{equation}
\noindent Here, $\Delta_S$ is the splitting of the register states (e.g. by an applied field), $\beta, \eta$ are chosen to respect the interaction symmetry at the injection points, and $g_L$, $g_R$ represent the interaction strength between the registers and the injection spins ($a$ and $b$) as shown in Fig.~2.  Transforming to Majorana operators yields $H_{int} =- \frac{\Delta_S}{2} (i \gamma^{3}_{L} \gamma^{0}_{L}  + i \gamma^{3}_{R} \gamma^{0}_{R}) + g_L \gamma^{1}_{L} \gamma^{1}_{a}  \gamma^{0}_{L} \gamma^{0}_{a} + g_R  \gamma^{1}_{R} \gamma^{1}_{b}  \gamma^{0}_{R} \gamma^{0}_{b}$, where, without loss of generality, we have chosen a $\sigma^{x} \sigma^{x}$ register-edge interaction. 

The existence of a dangling Majorana at the droplet edge is critical to enable spin-edge coupling. At the injection points (Fig.~2b), the register-edge coupling of equation (4), not only creates a fermionic excitation, but also introduces a dangling vortex (by flipping the $U_{i,j}$ corresponding to the adjacent dangling plaquette). Thus, in order to exploit the chiral fermion mode to transport spin-based quantum information, we will need to control the injection point (see Supplementary Information for additional details). Imperfections in such control will result in the spin-register coupling to additional nearby spins. However, since the nearest spins surrounding the injection point will not contain dangling Majoranas, these additional interactions will naturally gap out.  

Even in the presence of the additional interactions prescribed in equation (4), since $U_{L,a}$ and $U_{R,b}$ are conserved, the model remains exactly solvable. Expressed in terms of the eigenmodes of the unperturbed Hamiltonian in the ground state flux sector,
\begin{equation}
\begin{aligned}
&H_T = H + H_{int} = \sum_{k>0} \epsilon_k (c_k^{\dagger} c_k - \frac{1}{2}) \\
&+ \Delta_S ( c_{L}^{\dagger} c_{L} -\frac{1}{2})+ \Delta_S ( c_{R}^{\dagger} c_{R} -\frac{1}{2}) \\
& - g_L U_{L,a} (c_{L} + c_{L}^{\dagger})  \frac{i}{\sqrt{2}} (\sum_k Q^{*}_{k,a} c_k + \sum_k Q_{k,a} c_k^{\dagger})\\
&- g_R U_{R,b}(c_{R} + c_{R}^{\dagger})  \frac{i}{\sqrt{2}} (\sum_k Q^{*}_{k,b} c_k + \sum_k Q_{k,b} c_k^{\dagger}),\\
\end{aligned}
\end{equation}
\noindent where we have defined $c_{L,R}^{\dagger} =1/2( \gamma_{0}^{L,R} -i \gamma_{3}^{L,R})$ and $c_{L,R} = 1/2(\gamma_{0}^{L,R} +i \gamma_{3}^{L,R})$; in this language, the $\sigma^z$ spin state of the L(R) qubit is encoded in the occupation of the L(R) fermion mode.  The first term of the Hamiltonian characterizes the modes of the 2D droplet, the second and third term characterize the splitting associated with the spin-registers, while the final two terms capture the coupling between the registers and the dangling edge spins.  This Hamiltonian acts in the extended fermionic Hilbert space and returning to physical spin states requires gauge projection (see Supplementary Information for details). 

\subsubsection*{Topologically Protected State Transfer in the Dot and Droplet Regime}

The coupling between the register and the chiral edge mode can be analyzed in two distinct regimes: 1) the mesoscopic dot regime and 2) the macroscopic droplet regime. The distinction between these two regimes is best understood from a perspective of resolvability; in the dot regime, we consider the coupling to a small finite-size system, enabling energy resolution of the individual chiral edge modes. Thus, TPST is mediated by a single fermionic eigenmode of the system \cite{YJG10}. Meanwhile, in the droplet regime, we consider the coupling to a larger system, in which energy resolution at the single mode level would be extremely difficult. In this regime, we encode the spin register's quantum information in a traveling fermionic wave-packet.

In both the dot and droplet regimes, TPST relies on the coherent transfer of fermionic occupation from register $L$ to $R$.  In order for this to be well-defined, we choose $\Delta_S > 0$ and $g_L, g_R < \Delta_S$ so that the effective Dirac fermions, $c^\dagger_k$, are conserved. 
In the dot regime, TPST can be understood by tuning $\Delta_S$ to be resonant with a single edge mode, $\tilde{k}$, with energy $\epsilon_{\tilde{k}}$; so long as the coupling strength is weak enough to energetically resolve this mode, evolution is governed by the effective Hamiltonian, 
\begin{equation}
H_{eff} =  -\frac{i}{\sqrt{2}} g_L  Q^{*}_{\tilde{k},a} c_{L}^{\dagger} c_{\tilde{k}} -\frac{i}{\sqrt{2}} g_R Q^{*}_{\tilde{k},b} c_{R}^{\dagger} c_{\tilde{k}} + h.c.
\end{equation}
\noindent and hence, state transfer proceeds via resonant fermion tunneling, as depicted in Fig.~4a (for details see Supplementary Information) \cite{YJG10}. 
The timescale, $\tau$, required to achieve high fidelity state transfer depends only on the energy spacing between adjacent modes, $\Delta \epsilon \sim \kappa/\ell$, where $\ell$ is the system's linear dimension; to prevent the leakage of quantum information into off-resonant fermionic modes, $\tau \gtrsim \ell/\kappa$ \cite{YJG10}.

\begin{figure}
\centering
\includegraphics[width=3.4in]{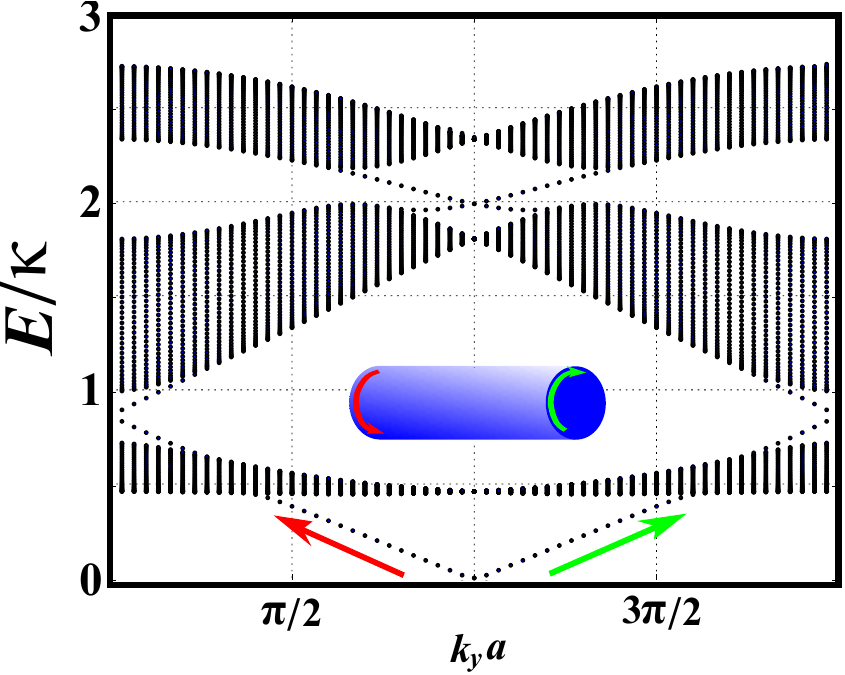}  
\caption{\label{fig:Fig3} {\bf Fermionic spectrum of the ground state flux sector $|$} The model is placed on a cylinder of circumference $61$ and width $40$ unit cells, with a zigzag edge oriented such that the $y$ direction is periodic \cite{Kohmoto07}. The chiral edge modes are clearly visible near $k_y a = \pi$ at energies below the bulk fermion gap $\Delta_b=0.46\kappa$. Numerical simulations also indicate the following values of the two vortex gaps: $0.14 \kappa$ (dodecagonal vortex) and $0.17 \kappa$ (triangular vortex).  The inset illustrates the oppositely propagating chiral edge mode (red and green) at each end of the cylinder. In a droplet, the flux sector with completely opposite flux in each plaquette would support edge modes traveling in a reversed orientation. }
\end{figure}

In the droplet regime, we encode the fermionic occupation into the presence/absence of a wavepacket traveling along the chiral edge, as illustrated in Fig.~4b \cite{Osborne04}. Upon tuning both spin-registers to an energy $\Delta_S$, the encoding can be performed by choosing $g_L(t)$ with the following time-dependence,
\begin{eqnarray}
g_L(t) = \frac{\sqrt{v}f(t)}{\sqrt{ \int_t^\infty d t' |f(t')|^2}},
\end{eqnarray}
\noindent where $f(t)$ characterizes the shape of the desired wave-packet and $v$ is the group velocity of the chiral mode. Subsequent retrieval can be similarly achieved by employing time-reversal symmetry to appropriate choose the shaping of $g_R(t)$ (see Supplementary Information for details). We note that such wavepacket encoding is in direct analogy to the storage and retrieval of photonic wavepackets \cite{Lukin00, Sherson06, Gorshkov07}. In contrast to the dot regime, the magnitude of the coupling strengths may be of order $\Delta_S$, which is independent of $\ell$. However, the time scale of TPST includes the wavepacket's propagation time, which depends on both the physical separation of the registers and the wavepacket group velocity.

\section*{Effects of  Imperfections, Disorder and Decoherence}

Having explicitly demonstrated TPST in an exactly solvable model, we now consider additional imperfections, disorder, temperature and decoherence. As the CSLB phase has a bulk gap and a topological invariant protecting its chiral edge mode, we expect the effective low energy fermion dynamics to be insensitive to small perturbations \cite{Kitaev06}. Furthermore, the chirality of the edge mode prevents localization and the Majorana nature of the edge fermions strongly suppresses the phase space for scattering, thereby limiting nonlinear corrections to the dispersion \cite{Bose07, YJG10, Christandl04, Clark05, Fitzsimons06}. In the following, we consider various classes of imperfections arising from local spin perturbations and coupling to a finite temperature bath; these result in: 1) vortex excitations, 2) finite Majorana lifetime and 3) dynamical decoherence.

\subsubsection*{Vortex Excitations}
At low temperatures $T$, there will be a dilute gas of bulk vortices, $N_v \sim n_p e^{-\Delta_v/T}$, where $n_p$ represents the total number of bulk plaquettes. Since a vortex excitation corresponds to a $\pi$ flux relative to the ground state, a circumambulating fermion acquires an additional phase of $N_v \pi$. Thus, the presence of vortices can have two relevant effects: 1) vortices within a localization length, $\xi \sim a$ (where $a$ is the lattice spacing), of the edge can scatter a traveling fermion and 2) an odd number of vortices induces a $\pi$-shift of the net phase \cite{Fendley09}. 

In addition to introducing bulk vortices, perturbations also generically lift the aforementioned degeneracy associated with dangling edge vortices. 
However, this will only affect the fidelity of TPST at the injection point, where one must ensure the existence of a single dangling edge Majorana.  Away from the injection point, three possibilities arise: First, zero energy dangling vortices are completely decoupled from the fermions and hence will be irrelevant for TPST. Second, low energy dangling vortices will scatter only minimally, since the interaction strength between the dangling Majoranas, and hence the hopping strength across the dangling link, will be extremely weak (see Supplementary Information for details).  Finally, much as in the bulk, the effect of high energy dangling vortices will be suppressed by their gap. 

As static effects, all of the aforementioned error contributions can be abrogated by the use of tomography; hence, it is crucial to effectively freeze out vortex fluctuations on the time scale of TPST, and this is most easily accomplished at temperatures which are small compared to $\Delta_v$. 

\subsubsection*{Finite Majorana Lifetime}

\begin{figure}
\centering
\includegraphics[width=3.4in]{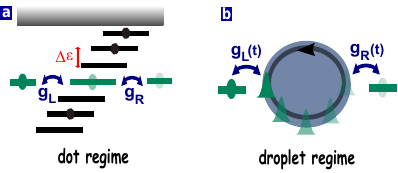}
\caption{\label{fig:Fig4} {\bf Regimes of TPST $|$} (a) Schematic representation of the dot regime wherein TPST becomes analogous to tunneling. In this mesoscopic dot regime, the coupling strength is kept weak enough to enable resolution of single edge modes. (b) Schematic representation of the droplet regime wherein TPST is achieved by mapping the quantum information from a spin-register onto a traveling fermionic wave-packet. The wave-packet is caught at the remote register, after which a two-qubit gate (see Supplementary Information for details) is performed before the information is returned (via a wave-packet) to the initial register. }
\end{figure}

Next. we consider the addition of generic perturbative local spin interactions, $H_p$, to the full Hamiltonian, $H_T$. Certain classes of perturbations leave the model exactly solvable; more generally however, if $H_p$ is longer ranged or does not respect the model's interaction symmetry, the gauge field acquires dynamics and the effective fermionic theory is no longer free. In order to understand these effects, we turn to a low-energy continuum theory of the Majorana edge (assuming that dangling vortex excitations are either decoupled or gapped out),
\begin{equation}
	H_e = v \int \frac{dp}{2\pi} p c^\dagger_p c_p = v \int dx\, \gamma(x) (i\partial) \gamma(x) ,
\end{equation}
\noindent where $c^\dagger_p = c_{-p}$ is the subset of $\{ c_k^{\dagger} \}$ in equation (3) which creates an edge excitation at momentum $p$ and where we have switched to a continuum normalization of the Majorana field, $\{\gamma(x),\gamma(y)\} = \delta(x-y)$ \cite{Fendley09}.

The introduction of interactions induces decay of the quasiparticle excitations $c^\dagger_p$. This quasiparticle lifetime limits the size of the droplet around which coherent excitations may be sent. The leading order symmetry-allowed interaction is of the form \cite{Fu09}
\begin{equation}
H'_{e} = \lambda \int dx\, \gamma(x) (i\partial)  \gamma(x) (i\partial)^2 \gamma(x) (i\partial)^3  \gamma(x),
\end{equation}
\noindent where $\lambda$ characterizes the strength of the interaction.  We estimate the decay rate $\Gamma_p^{int}$ of a single quasiparticle excitation using Fermi's golden rule (see Supplementary Information for details). In the low temperature limit ($\epsilon_p \gg k_BT$), 
\begin{equation}
\Gamma_p^{int} \sim \frac{\lambda^2 p^{13}}{v} + \frac{\lambda^2 p^{11} T^2}{v} + \mathcal{O}(T^4).
\end{equation}
To relate $\Gamma_p^{int}$ to the microscopic model parameters, we consider generic vortex-inducing local spin perturbations of strength $\kappa'$, which yield $\lambda \sim \kappa (\frac{\kappa'}{\kappa})^2 a^7$ in second order perturbation theory. Substituting into equation (10) allows us to re-express the zero temperature decay rate as $\Gamma_p^{int} \sim \frac{\kappa^2}{\Delta_S}  (\frac{\kappa'}{\kappa})^4 (a p)^{14}$, where $\Delta_S = vp$ is the energy of the injected TPST fermion. The surprisingly strong dependence on momenta suggests that quasiparticle decay can safely be neglected so long as $p < 1/a$.


\subsubsection*{Dynamical Decoherence}

Finally, we consider dynamical decoherence due to weak coupling with a low temperature phonon bath, which induces additional decay $\Gamma_p^{dec}$ of the fermion involved in TPST. We assume that the bath couples to local spin operators $\sigma^\alpha_i$ and that its effect is characterized by its noise spectral density, $\Gamma_p^{dec} \sim S(\omega)$ \cite{Taylor06}. In the bulk, each such operator creates a pair of vortices (Fig.~2b) in addition to creating or destroying a Majorana quasiparticle. As the fermionic edge modes are exponentially localized, the contribution of this process to the decay rate is suppressed by $e^{-d/\xi}$, where $d$ is the distance from site $i$ to the edge. Moreover, there is an additional energy suppression from $S(\omega_0) \sim e^{-\omega_0/k_BT}$ where $\omega_0 = 2\Delta_v$ is the energy cost of creating a pair of vortices.

\begin{figure}
\centering
\includegraphics[width=3.4in]{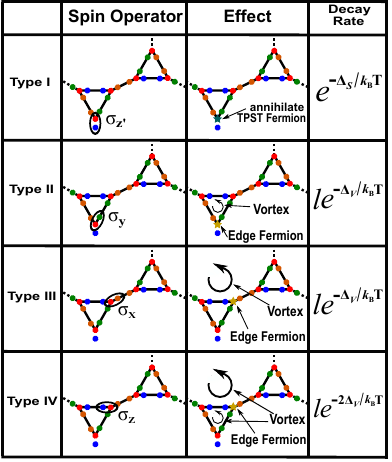}
\caption{\label{fig:Fig5} {\bf Schematic representation of the various forms of edge decoherence  $|$}  Type I spin operators correspond to non-vortex inducing decoherence and can affect TPST only by annihilating the TPST fermion (teal star), a process suppressed by $e^{-\Delta_S/k_BT}$, where $\Delta_S$ represents the detuning of the spin registers and hence also the energy of the injected quasiparticle. Type II-IV spin operators create vortices (circular arrow) in edge plaquettes. Once a vortex is created at any edge site, it can scatter the traveling TPST fermion, leading to the decoherence being enhanced by $\ell$ (the droplet's linear dimension). In addition to creating vortices, Type II-IV spin operators also create un-gapped edge fermions (gold star), which we assume does not affect TPST since quasiparticle interactions have been shown to be extremely weak.}
\end{figure}

This brings us to the primary decoherence effect: edge noise. There are two types of spin operators acting on the lattice edge: 1) those which only create or destroy an edge fermion (type I) and 2) those which also introduce vortices (type II-IV), as shown in Fig.~5. Type I spin operators can only induce decay if they directly annihilate the injected TPST edge fermion, a process costing energy $\Delta_S$. By contrast, once a vortex is created at any edge plaquette it can scatter the traveling TPST fermion, implying that the associated decoherence is enhanced by a factor of $\ell$, as depicted in Fig.~5. Thus, the total TPST decay rate induced by edge noise is,
\begin{equation}
	\Gamma_p^{dec} \sim e^{-\Delta_S/k_BT} + \ell  e^{-\Delta_V/k_BT}.
\end{equation}
Strikingly, the sources of decoherence in TPST are exponentially suppressed in temperature and thus can be controlled \cite{YJG10}.

The above analysis generalizes to other types of noise sources. Of particular relevance in the context of solid-state spin systems are nuclear spin baths, in which $S(\omega) \sim \frac{1}{\omega^2+1/t_c^2}$, where $t_c$ is the bath's correlation time. In this model, the Arrhenius-type energy suppressions of equation (11) becomes $\Gamma_p^{dec} \sim 1/\Delta_v^2$ if $\Delta_v \gg 1/t_c$ \cite{Taylor06}. 

\section*{Experimental Realizations and Outlook}

The search for novel topological phases represents one of the most exciting challenges in many-body physics; indeed, this challenge has led to a widespread effort to experimentally identify or engineer systems exhibiting exotic topological order. One of the prototypes of such order is provided by the CSLB phase of the Kitaev honeycomb model; while such chiral spin liquid phases have yet to be experimentally implemented, several realistic approaches toward their realization have been envisioned.  


The realization of a honeycomb lattice, an essential component of implementing the Kitaev gapped B phase, is currently being considered in systems ranging from  
ultra-cold atoms \cite{Duan03, Zhang07, Sugawa10} and polar molecules \cite{Micheli06, Buchler07} to superconducting lattices \cite{Gladchenko09, You10} and dipolar-coupled electronic spin arrays \cite{YJG11,Spinicelli11}. While engineering a macroscopic honeycomb droplet remains a daunting challenge, recent experiments have demonstrated the ability to control mesoscopic ensembles containing tens of qubits \cite{Buluta09, Gao10, Johnson11, Islam11, Monz11}. Despite these remarkable advances, such mesoscopic systems are insufficient in size to support the existence of several well-separated quasiparticles, a crucial prerequisite to demonstrate the non-abelian braiding essential for topological quantum computing  \cite{Kitaev06, DasSarma06, Nayak08}. However, these smaller systems represent ideal candidates to demonstrate topologically protected state transfer and hence, the existence of a chiral fermion edge - another hallmark of the CSLB phase.

Moreover, our proposed technique can also be used to directly characterize spin liquid states via passive spectroscopy of the droplet edge. By observing the splitting-dependent relaxation of the spin-qubit probe, one could map the energy spacing between the chiral edge-modes. In addition, asymmetries in correlation measurements provide a direct indication of chirality. In this case, by gradually altering the physical distance separating two spin-qubit probes, one could characterize the timescale of incoherent interactions between the remote registers.  Asymmetry in this timescale, dependent on the direction in which the qubits are separated provides a strong indication of the existence of a chiral edge and would enable direct evaluation of the velocity associated with the edge dispersion. Alternatively, one could also imagine holding the spin qubits fixed and characterizing asymmetries associated with $L$-to-$R$ versus $R$-to-$L$ TPST. These considerations imply that solid-state magnetic spin probes can provide a potential tool for exploring the properties of natural spin liquid candidates in both organic and inorganic insulators \cite{Hermele09, Lee08, Balents10, Machida10}.

Finally, our technique suggests a new avenue for a hybrid solid-state quantum computing architecture. In particular, while solid-state spin systems enable the realization of long-coherence-time quantum memories and high-fidelity quantum registers, scaling these individual components up to a local area quantum network remains a critical challenge \cite{Dutt07, Balasubramanian09, Neumann10b,YJG11,Weber10,YJG10}. Thus, we envision a hybrid architecture in which conventional solid-state spin qubits are connected by topologically protected channels. In this scenario, dynamical decoupling of spins within an engineered CSLB droplet generates a lattice of mesoscopic CSLB islands around which individual qubits reside \cite{Lange10}.  The edge-modes of these island regions act as quantum routers, ferrying quantum information between remote spin registers. Since spin-qubits are naturally well-separated in such a hybrid architecture, the individual addressing and direct control of single qubits is greatly simplified \cite{Maurer09}. Furthermore, such an architecture suggests a perspective in which control fields can reshape CSLB islands and thereby dynamically reconfigure network connectivities.

\section*{Acknowledgements}

We gratefully acknowledge conversations with T. Kitagawa, S. Bennett, P. Maurer, E. Altman, E. Demler, S. Sachdev, M. Freedman, and J. Preskill. This work was supported, in part, by the NSF, DOE (FG02-97ER25308), CUA, DARPA, AFOSR MURI, NIST, Lawrence Golub Fellowship, Lee A. DuBridge Foundation and the Sherman Fairchild Foundation. H.W. was supported by the National Science Foundation through a grant for the Institute for Theoretical Atomic, Molecular and Optical Physics at Harvard University and the Smithsonian Astrophysical Observatory and by a fellowship within the Postdoc Program of the German Academic Exchange Service (DAAD).

\end{document}


\date{\today}

\title{SUPPLEMENTARY INFORMATION \\
Topologically Protected State Transfer in a Chiral Spin Liquid}

\author{N. Y. Yao$^{1\dagger*}$, C. R. Laumann$^{1,2\dagger}$, A. V. Gorshkov$^{3}$, H. Weimer$^{1,2}$, L. Jiang$^{3}$, J. I. Cirac$^{4}$, P. Zoller$^{5}$, M. D. Lukin$^{1}$}

\affiliation{$^{1}$Physics Department, Harvard University, Cambridge, MA 02138}
\affiliation{$^{2}$ITAMP, Harvard-Smithsonian Center for Astrophysics, Cambridge, MA 02138}
\affiliation{$^{3}$Institute for Quantum Information, California Institute of Technology, Pasadena, CA 91125}
\affiliation{$^{4}$Max-Planck-Institut fur Quantenoptik, Hans-Kopfermann-Strase 1, Garching, D-85748, Germany}
\affiliation{$^{5}$Institute for Quantum Optics and Quantum Information of the Austrian Academy of Sciences, A-6020 Innsbruck, Austria}
\affiliation{$^{\dagger}$These authors contributed equally to this work}
\affiliation{$^{*}$e-mail: nyao@fas.harvard.edu}
\maketitle

\section{Gauge Projection to the Spin Subspace} 
\label{sec:gauge_projection}

\noindent Having demonstrated topologically protected state transfer by working in the extended Hilbert space in the main text, here, we consider the gauge projection back to the physical subspace. We ultimately illustrate the SWAP gate associated with TPST in the language of physical spin states. Recall that the 4-Majorana representation of the spin algebra $\sigma^\alpha = i \gamma^\alpha \gamma^0$ (for spins $\alpha \in \{x,y,z \}$ and correspondingly for Majoranas $\alpha \in \{1,2,3 \}$) only holds under the constraint that $D = \gamma^1 \gamma^2 \gamma^3 \gamma^0 = 1$ \cite{Kitaev06}. This defines the two dimensional physical subspace of the four dimensional Hilbert space associated to four Majorana fermions. More generally, for $N$ spins $\sigma^\alpha_i$, the $2^N$ dimensional physical space is defined by the $N$ constraints $D_i = 1$ in the $2^{2N}$ dimensional extended Hilbert space. We may impose these constraints by using the physical (gauge) projector
\begin{equation}
	P = \prod_i \frac{1+D_i}{2} \tag{S1}\\
\end{equation}
which simply annihilates any state not satisfying the local constraint.

\vspace{5mm}

\noindent The gauge projector $P$ has several important properties. It commutes with any physical operator built out of spin operators: $[P, \sigma^\alpha_i] = 0$.  Furthermore, $P$ absorbs gauge transformations $D_i$, with $PD_i = D_i P = P$. This gives us the freedom to understand the evolution of physical states $\ket{\psi}$ under any physical operation $\mathcal{O}$, which is a function of spin operators $\sigma^\alpha_i$, by lifting this operation to the extended Hilbert space $\mathcal{O^\gamma} = \mathcal{O}(i \gamma^\alpha_i \gamma^\alpha_0)$, applying any gauge transformations $D_i$ which may simplify the analysis, and projecting back only at the end. That is, if we can find a state $\ket{\psi^\gamma}$ in the extended Hilbert space such that $\ket{\psi} = P \ket{\psi^\gamma}$ then $\mathcal{O} \ket{\psi} = P \mathcal{O^\gamma} \ket{\psi^\gamma}$. In particular, if $\ket{\psi^\gamma}$ is an eigenstate of $\mathcal{O}^\gamma$, the physical state $\ket{\psi}$ is an eigenstate of $\mathcal{O}$. This is particularly useful for identifying energy and spin eigenstates in the extended Hilbert space. One must always ensure that $P \ket{\psi^\gamma} \ne 0$. 

\subsection{Decorated Honeycomb Lattice} 
\label{sec:the_main_model}

\noindent We now focus on the decorated honeycomb lattice model along with two additional spin registers as in Fig. 2b. of the main text. The full Hamiltonian, $H_T = H_0 + H_L +H_R + H_{int}$ is composed of
\begin{align}
H_0 &= \frac{1}{2}\sum_{\genfrac{}{}{0pt}{}{x,x'}{links}}  \sigma^{x}_{i} \sigma^{x}_{j} + \frac{1}{2}\sum_{\genfrac{}{}{0pt}{}{y,y'}{links}}  \sigma^{y}_{i} \sigma^{y}_{j} + \frac{1}{2}\sum_{\genfrac{}{}{0pt}{}{z,z'}{links}}  \sigma^{z}_{i} \sigma^{z}_{j} \nonumber \\ 
H_{L/R}  &= -\frac{\Delta_S}{2} \sigma^z_{L} - \frac{\Delta_S}{2} \sigma^z_{R}, \hspace{3mm} H_{int} = g_L \sigma^x_L \sigma^x_a + g_R \sigma^x_R \sigma^x_b    \tag{S2}  \nonumber
\end{align}
where we have chosen units in which $\kappa = 1$. The extension of the Hamiltonian to the Majorana
Hilbert space results in a model of Majorana fermions $\gamma^0_i$ coupled to a static
$\mathbb{Z}_2$ gauge field $\hat{U}_{i,j}$ residing on the lattice links \cite{Yao07}:
\begin{align}
	H^{\gamma}_0 & = \frac{i}{4} \sum_{i j} \hat{U}_{i,j} \gamma^0_{i} \gamma^0_j \nonumber  \tag{S3}  \\ 
	H^{\gamma}_{L/R} & = -\frac{\Delta_S}{2} i \gamma^3_L \gamma^0_L - \frac{\Delta_S}{2} i \gamma^3_R \gamma^0_R \nonumber  \tag{S4}  \\ 
	H^\gamma_{int} & = - ig_L \hat{U}_{L,a} \gamma^0_L \gamma^0_a - ig_R \hat{U}_{R,b} \gamma^0_R \gamma^0_b \nonumber  \tag{S5}  
\end{align}
where $\hat{U}_{i,j} = i \gamma^\alpha_i \gamma^\alpha_j$ and $\alpha = x,y,z$ is the link type of $\langle ij \rangle$, or $\hat{U}_{i,j} = 0$ if $i$ and $j$ are not connected. We extend the definition of the gauge field $\hat{U}_{i,j}$ to the paired dangling edge spins on the boundary as in Fig. S1, and to $\hat{U}_{L,R} =  i \gamma^2_L \gamma^2_R$. With this choice of pairings, all $\hat{U}_{i,j}$ are conserved by the total Hamiltonian and thus time evolution may be understood in each $\hat{U}_{i,j}$ sector. We label the sectors of $\hat{U}_{i,j}$ by field configurations $\{U_{i,j} = \pm 1\}$. Due to the antisymmetry of $U_{i,j}$, there is some subtlety in correctly labeling sectors: in all our formulae, we take $ij$ to be oriented according to the arrows in Fig. 2b of the main text. Thus, $U_{i.j} = 1$ corresponds to a ground state gauge sector. Finally, we define $c_{L/R} = \frac{1}{2}(\gamma^0_{L/R} + i \gamma^3_{L/R})$ as in the main text so that we may think of $H^\gamma_{L/R}$ as measuring the occupation of left and right register fermions. 

\begin{figure}[t]
  \begin{center}
   \includegraphics[width=7in]{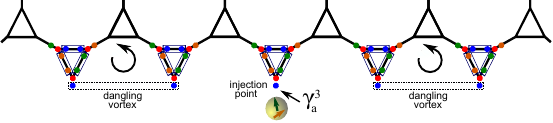}
  \end{center}
    \begin{flushleft}
\noindent FIG.~S1: \textbf{Dangling Majoranas and Vortices $|$} A schematic microscopic edge of the decorated honeycomb with the injection point explicitly labeled. Other pairs of dangling edge spins are paired into decoupled dangling vortices. It is these dangling vortices which lead to the large degeneracy found in the exactly solvable model.
\end{flushleft}
\end{figure}

\vspace{5mm}

\noindent The $U_{i,j}$ are gauge dependent quantities as $\{D_i, \hat{U}_{i,j}\} = 0$, but the net flux
around any closed loop $w(C) = \prod_{ij \in C} \hat{U}_{i,j}$ is gauge invariant; thus $w(C)$ is physical
and conserved \cite{Kitaev06}. The extended Hilbert space may be divided into conserved \emph{gauge} sectors
while the physical Hilbert space splits into conserved \emph{flux} sectors after projection. As
usual, we say that any plaquette $P$ such that $w(\partial P) = -1$ contains a vortex; here, we
additionally extend this definition of vortices to include the dangling plaquettes defined by the
$U_{i,j}$ links between dangling edges, as shown in Fig. S1. These \emph{dangling vortices} are completely
decoupled from the fermions and lead to a large degeneracy of the model with open boundaries
($2^{N_e/4}$ where $N_e$ is the number of dangling edge spins).


\subsection{Spin states on the Decorated Honeycomb} 
\label{sec:spin_states}

\noindent Let us consider the physical ground state of the system in the absence of interaction $g$ between the registers and the decorated honeycomb. The spin registers both point up, disentangled from the rest of the system, while the lattice spins sit in their collective ground state: $\ket{\u}_L\ket{GS}_{0}\ket{\u}_R$. We seek a reference ground state $\ket{\Omega}$ in a fixed gauge sector of the extended Hilbert space such that $\ket{\u}_L\ket{GS}_{0}\ket{\u}_R = P \ket{\Omega}$ up to normalization. We choose $U_{i,j}=+1$ (\emph{i.e.} the flux configuration of the ground state sector as described in the main text) and we choose $\ket{\Omega}$ to be annihilated by  $c_L$, $c_R$ and $c_k$ for $k>0$ of $H^\gamma_0(U) + H^\gamma_{L/R}$. This state is, by construction, a lowest energy eigenstate of the system, but it may not survive projection. The norm of $P\ket{\Omega}$ is
$\bra{\Omega}PP\ket{\Omega} = \frac{1}{2^N}\bra{\Omega}1+\prod_i D_i \ket{\Omega}$ where we have exploited the orthogonality of states with different gauge configurations $U_{i,j}$. The product over all gauge transformations $D_i$ measures the product of all $U_{i,j}$ and the parity of the $\gamma^0_1 \gamma^0_2 \cdots \gamma^0_N$ fermionic state.  Thus, by flipping the choice of ${U}_{L,R}$ in $\ket{\Omega}$, we may guarantee that $P\ket{\Omega}$ survives projection. In general, in any fixed gauge sector related to our reference sector by an even (odd) number of flipped $U_{i,j}$'s, $P$ will annihilate states with odd (even) fermionic parity.  

\vspace{5mm}

\noindent We now construct an explicit representation of the 4 possible register states coupled to the intermediate ground state $\ket{GS}_0$ by acting with $\sigma^{x}_{L/R}$:
\begin{align}
\label{eq:phys_spin_states}
	\ket{\u}_L\ket{GS}_{0}\ket{\u}_R &= P \ket{\Omega} \nonumber  \tag{S6}  \\
	\ket{\d}_L\ket{GS}_{0}\ket{\u}_R &= -i P c^\dagger_L \gamma^1_L \ket{\Omega} \nonumber\\
	\ket{\u}_L\ket{GS}_{0}\ket{\d}_R &= -i P c^\dagger_R \gamma^1_R \ket{\Omega} \nonumber\\
	\ket{\d}_L\ket{GS}_{0}\ket{\d}_R &= - i P c^\dagger_L c^\dagger_R \ket{\Omega} \nonumber   
\end{align}
where in the last line we have used $U_{L,R} = 1$ acting on $\ket{\Omega}$. More generally, degenerate and/or low energy states may be found in either the same flux sector with (pairs of) extra edge fermions, \emph{e.g.} $Pc^\dagger_{p}c^\dagger_{p'}\ket{\Omega}$ or in degenerate flux sectors (containing dangling vortices), \emph{e.g.} $Pc^\dagger_k \gamma^{\alpha}_{i}\ket{\Omega}$ where $\gamma^\alpha_i$ is a dangling edge Majorana. We note that since $H_0^\gamma(U)$ does not depend on the dangling edge $U_{i,j}$'s, neither do the fermionic eigenmodes $c_k$ in these degenerate sectors nor the fermionic vacuum.

\subsection{SWAP Gate in the Physical Subspace} 
\label{sec:state_transfer}

\noindent Let us now consider time evolution $\UU(t)$ in the presence of the coupling $H^\gamma_{int}$.
The gauge field $U_{i,j}$ remains conserved and the time evolution of the Majorana field $\gamma^0_i$ within each gauge sector is that of noninteracting fermions. The full Hamiltonian in  our chosen ground state gauge sector is given by equation (5) of the main text. In general, $\epsilon_k$, $Q_{k,a}$, and $c_k$ depend on the gauge field but not on the dangling pieces of it, so the following analysis applies identically in each sector containing  dangling vortices so long as the gauge is chosen the same way in the bulk and on $U_{L,a}$, $U_{R,b}$. 
Assuming that $g \ll \Delta_S$, we may use the secular approximation to eliminate the $c$-fermion number non-conserving terms in equation (5), 
\begin{align}
	H^\gamma(U) &\approx \sum_{k>0} \epsilon_k (c_k^{\dagger} c_k - \frac{1}{2}) 
	+ \Delta_S ( c_{L}^{\dagger} c_{L} -\frac{1}{2})+ \Delta_S ( c_{R}^{\dagger} c_{R} -\frac{1}{2}) \nonumber  \tag{S7}  \\
	& -  \frac{i}{\sqrt{2}} g_L (c_{L}^{\dagger} \sum_k Q^{*}_{k,a} c_k + c_{L} \sum_k Q_{k,a} c_k^{\dagger}) \nonumber\\
	&-  \frac{i}{\sqrt{2}} g_R (c_{R}^{\dagger}\sum_k Q^{*}_{k,b} c_k + c_{R}\sum_k Q_{k,b} c_k^{\dagger}).	\nonumber
\end{align}
This Hamiltonian leaves the $c$-fermion vacuum $\ket{\Omega}$ invariant and evolves the modes as usual non-interacting Dirac fermions:
$\UU c_{k_1}^{\dagger}c_{k_2}^{\dagger} \cdots c_{k_m}^{\dagger} | \Omega \rangle =  c_{k_1(t)}^{\dagger}c_{k_2(t)}^{\dagger} \cdots c_{k_m(t)}^{\dagger} | \Omega \rangle$, where  $k_i(t)$ denotes the time evolved wavefunction of the $k_i$ mode according to the single particle Schr\"odinger equation. We note that the most general Majorana evolution would mix the $c^\dagger$ and $c$ modes and accordingly the instantaneous $c$-vacuum would evolve in time.

\begin{figure}[t]
  \begin{center}
   \includegraphics[width=4.5in]{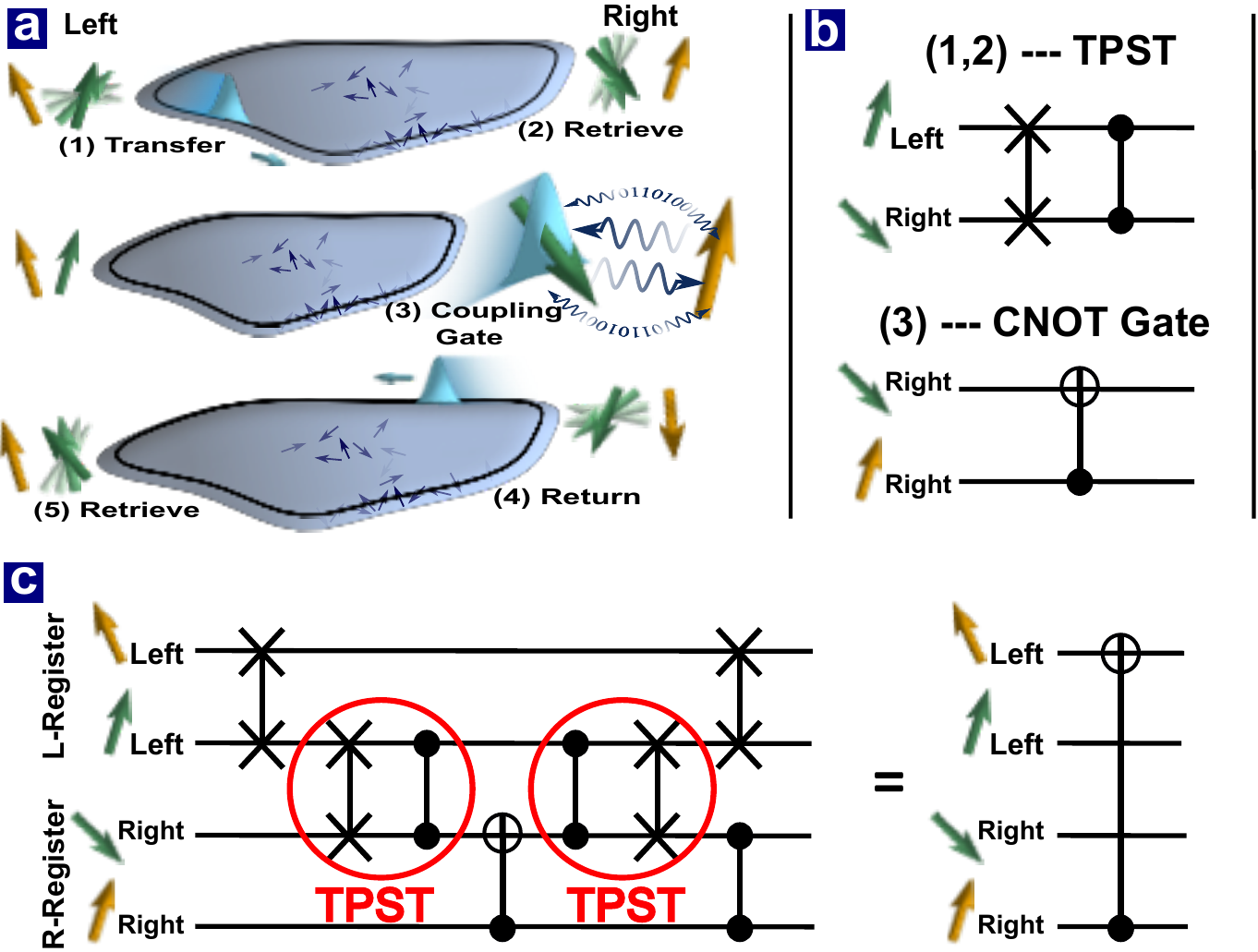}
  \end{center}
    \begin{flushleft}
\noindent FIG.~S2: \textbf{ Schematic representation of TPST and a remote CNOT gate $|$} (a) The quantum registers $L$ and $R$ each contain two spins with the gold spin corresponding to the memory qubit. A single step of evolution SWAPs the information between the green spins of the left and right register. However, in addition to the SWAP gate, it also creates entanglement in the form of a controlled phase gate between these spins. To perform a remote CNOT gate between the memory qubits, we first perform an intra-register operation to SWAP the quantum information between the green and gold qubit of the left spin. (b) Next, the first step of TPST is performed (corresponding to a SWAP gate and a controlled phase gate). Afterwards, an intra-register CNOT gate between the green and gold qubit of the right register is performed. (c) The second step of TPST is then performed to return the information to the left register. Finally, intra-register gates are performed to yield a remote CNOT between the memory qubits. This enables universal computation. 
\end{flushleft}
\end{figure}

\vspace{5mm}

\noindent To enable state transfer, we now tune $\Delta_S = \epsilon_{\tilde{k}}$ for an edge mode $\tilde{k}$. In the dot regime, we further require $|g_{L} Q_{\tilde{k},a}| , |g_{R} Q_{\tilde{k},b}| \ll | \epsilon_{\tilde{k}} - \epsilon_{\tilde{k}\pm1}|$. This condition enables single-mode resolution of the edge eigenmodes and state transfer proceeds by resonant fermionic tunneling in an effective three mode model (dropping constants and the uninvolved modes) \cite{Yao11}:
\begin{align}
H_{eff} &=  -\frac{i}{\sqrt{2}} g_L  Q^{*}_{\tilde{k},a} c_{L}^{\dagger} c_{\tilde{k}} -
\frac{i}{\sqrt{2}} g_R Q^{*}_{\tilde{k},b} c_{R}^{\dagger} c_{\tilde{k}} + h.c. \nonumber  \tag{S8}  
\end{align}
Since the individual quantum registers are fully controllable, we tune $g_L$ and $g_R$ to ensure that the effective tunneling rate $t_{\tilde{k}}= |  \frac{g_L}{\sqrt{2}} Q_{\tilde{k},a} |= | \frac{g_R}{\sqrt{2}} Q_{\tilde{k},b} |$ between the modes is equivalent. Re-expressing $-i Q^{*}_{\tilde{k},a}= e^{i \phi_{\tilde{k},a}} |Q_{\tilde{k},a}|$ and $-i Q^{*}_{\tilde{k},b} = e^{i \phi_{\tilde{k},b}}|Q_{\tilde{k},b} |$, subsequent evolution under $H_{eff}$ for a time $\tau = \frac{\pi}{\sqrt{2} t_{\tilde{k}}}$ results in mode evolution, 
\begin{align}
	c^\dagger_L &\longrightarrow - e^{-i\phi} c^\dagger_R   \nonumber  \tag{S9}  \\
	c^\dagger_{\tilde{k}} &\longrightarrow -  c^\dagger_{\tilde{k}} \nonumber \\
	c^\dagger_R &\longrightarrow - e^{i\phi} c^\dagger_L  \nonumber
\end{align}
where $\phi = \phi_{\tilde{k},a}-\phi_{\tilde{k},b}$. Using these relations to evolve the states from equation \eqref{eq:phys_spin_states}, we find
\begin{align}
	\ket{\u}_L\ket{GS}_{0}\ket{\u}_R &\longrightarrow \ket{\u}_L\ket{GS'}_{0}\ket{\u}_R \nonumber   \tag{S10}  \\
	\ket{\d}_L\ket{GS}_{0}\ket{\u}_R &\longrightarrow  - i e^{-i\phi}   \ket{\u}_L\ket{GS'}_{0}\ket{\d}_R \nonumber \\
	\ket{\u}_L\ket{GS}_{0}\ket{\d}_R &\longrightarrow  i e^{i\phi}   \ket{\d}_L\ket{GS'}_{0}\ket{\u}_R \nonumber \\
	\ket{\d}_L\ket{GS}_{0}\ket{\d}_R &\longrightarrow  - \ket{\d}_L\ket{GS'}_{0}\ket{\d}_R. \nonumber
\end{align}
up to known dynamical phases. The time evolution presented in equation (S10) generates our desired SWAP gate in addition to a controlled phase gate between the register modes (up to single qubit rotations). Here $\ket{GS'}_0$ indicates a state which evolves from $\ket{GS}_0$ \emph{independent} of the state of the two register qubits. As depicted in Fig.~S2, in combination with intra-register manipulations, the gate described by equation (S10) enables universal computation between the memory qubits of the remote spin registers \cite{Yao11}.

\vspace{5mm}

\noindent This schematic evolution holds identically for \emph{any} initial state of the intermediate system $\ket{GS}_0$ containing extra fermions or dangling vortices, since such states may be represented in a gauge sector where all bulk $U_{i,j} = 1$.  Furthermore, in flux sectors in which there are an \emph{even} number of bulk vortices, it is possible to choose a gauge in which $U_{i,j}=1$ for all links near the edge. The evolution proceeds nearly identically in this case as well. On the other hand, in flux sectors where there are an \emph{odd} number of bulk vortices, the energy of the edge modes is shifted by $\sim \kappa/L$ implying that the spin registers are off-resonant. This can be corrected for through tomography and subsequent retuning \cite{Burgarth09}. 

\section{Shaping the Traveling Fermionic Wavepacket}

\noindent Here, we describe the shaping of the fermionic wavepacket in the droplet regime of TPST. The edge mode energies $\epsilon_k$ of a finite-sized droplet are split at order $1/\ell$. As discussed in the main text, since single mode energy resolution becomes impossible in the macroscopic limit, we encode the spin register's quantum information into a fermionic wavepacket traveling along the chiral edge of the 2D droplet. This requires the shaping of $g_L(t)$ and $g_R(t)$ in order to ensure the sending and receiving of the packet. Let us first consider the shaping of the initial wavepacket at the left register, so $g_R(t)=0$. As described in Sec.~IC, here, it is sufficient for us to consider the single particle problem since the modes evolve as usual non-interacting Dirac fermions.  By tuning $\Delta_S$ to an energy in the middle of the edge dispersion and restricting $|g_L| \ll \Delta_S$, we have (assuming a plane wave description of the low energy chiral edge modes) 
\begin{align}
H_{wp} = \sum_{k} E_k |k\rangle \langle k| + \frac{g_L}{\sqrt{l}} \sum_k( |k\rangle \langle L | + |L \rangle \langle k |), \nonumber   \tag{S11}
\end{align}
\noindent where $|k\rangle$ is the edge mode with momentum $k$, we have absorbed all numerical factors into $g_L$ and $E_k=vk$ is shifted by $\Delta_S$ (here, we have correspondingly shifted the definition of zero energy and the indexing of $k$ to begin at the state with energy $\Delta_S$). We choose this notation for the Hamiltonian to be consistent with the literature regarding photonic wavepacket storage and retrieval, where an analogous problem is solved \cite{Lukin00,  Sherson06,Gorshkov07}; thus, in this section, $c_i$, rather than being fermionic operators, will represent the amplitude of the $|i\rangle$ mode.   Initially, we consider a state $|\psi \rangle$ whose amplitude is fully localized on the left spin register, $| \psi \rangle = c_L |L \rangle + \sum c_k |k \rangle$, where $c_L(t=0)=1$ and $c_k(t=0)=0$. After making a continuum approximation in both position and momentum, we formally solve the Schr\"odinger equation to obtain $\dot c_L(t) = - \frac{1}{2v} |g_L(t)|^2  c_L(t)$, yielding $c_L(t) = e^{h(t)}$ where $h(t) =  \frac{1}{2 v} \int_0^t d t' |g_L(t')|^2$. Substituting this result into the formal solution of $c_k(t)$ yields
\begin{align}
c_k(t)=-i \int_0^t d t' e^{-ivk(t-t')}\frac{1}{\sqrt{l}}g_L(t') e^{-h(t')}. \nonumber   \tag{S12}
\end{align}
\noindent Thus, the shape of the outgoing wavepacket is 
\begin{align}
c(x,t) &=  \frac{1}{\sqrt{l}} \sum_k e^{ikx} c_k(t) \approx \frac{\sqrt{l}}{2 \pi} \int d k e^{i k x} c_k(t) =  \frac{-i}{2 \pi} \int d k e^{i k x}  \int_0^t d t' e^{-ivk(t-t')}g_L(t') e^{-h(t')} \nonumber \tag{S13} \\
&=  - i   \int_0^t d t' \delta(x - v (t-t')) g_L(t') e^{- h(t')} =  - i \frac{1}{v} g_L(t-x/v) e^{-h(t-x/v)} \theta(t-x/v). \nonumber
\end{align}
where $\theta$ is the Heaviside step function and we have assumed linear dispersion with group velocity $v$. Here, we note that in converting from a $k$ sum to an integral, we have assumed that the amplitude on both $k<0$ and bulk modes will be negligible since $|g_L| \ll \Delta_S$. As previously discussed, this assumption is crucial to ensure that the vacuum does not undergo time evolution. 

\vspace{5mm}

\noindent It is natural to think of the wave-packet in the time domain and evaluate $c(x,t)$ at $x = 0$. Thus, the solution to the problem of shaping any desired wavepacket, $f(t)$, simplifies to deriving the requisite $g_L(t)$ control function that satisfies $\frac{1}{v}g_L(t) e^{-h(t)} = f(t)$ where $h(t) =  \frac{1}{2 v} \int_0^t d t' |g_L(t')|^2$; such a solution then yields,
\begin{align}
g_L(t) = \frac{\sqrt{v}f(t)}{\sqrt{ \int_t^\infty d t' |f(t')|^2}}. \nonumber \tag{S14} 
\end{align}
\noindent The subsequent retrieval of the wave-packet at the location of the right spin register can be understood by using time-reversal; indeed, the control function $g_R(t)$ should be the time-reversed form of the control used to generate the time-reversed form of the sent wavepacket \cite{Gorshkov07}. While, for simplicity, we have considered $g_L, g_R \in \mathbb{R}$ above, generalizing to complex $g_{L/R}$ can easily be achieved, for example, by employing a $\Lambda$-configuration spin register \cite{Gorshkov07}. One possible implementation of such a register is naturally provided by defect centers in diamond; in particular, Nitrogen-Vacancy color centers, which represent promising high-fidelity two qubit quantum registers, harbor a spin triplet ($S=1$) electronic ground state \cite{Dutt07, Neumann10b}.

\section{The Edge --- Injection and Interactions}
\noindent There are two types of excitations on the edge: Majorana fermions and $\mathbb{Z}_2$ vortices \cite{Yao07}. The topology of the bulk guarantees the existence of  gapless chiral Majorana edge-modes, as described in a low energy theory by,
\begin{equation}
\label{eq:ham_edge_maj}
H_e = v\int dx \gamma(x) (i\partial) \gamma(x) \nonumber \tag{S31} 
\end{equation}
\noindent where $\gamma(x)$ is a continuum Majorana field and $v$ is the group velocity \cite{Fendley09}. Generically all vortices are gapped; however, details of the lattice edge can lead to the existence of decoupled and/or low energy vortices. While the Hamiltonian \eqref{eq:ham_edge_maj} does not capture these degrees of freedom, the presence of these additional states in the low energy Hilbert space cannot be ignored. Indeed, the degeneracy of the exactly solved model follows from the presence of zero-energy dangling vortices formed by pairs of dangling Majoranas, as depicted in Fig. S1. Away from the injection point: 1) zero energy vortices are decoupled and hence irrelevant to TPST, 2) low energy vortices scatter only weakly, and 3) high energy vortices are suppressed by temperature. The presence of a low-energy vortex degree of freedom at the injection point is critical to enable spin-edge coupling, which occurs at a dangling spin. Crucially, this dangling spin contains a decoupled Majorana operator $\gamma_a^3$ (dangling Majorana), as shown in Fig. S1.  Keeping track of this mode in the low-energy Hilbert space allows us to couple as follows, 
\begin{equation}
H_c = v\int dx \gamma(x) (i\partial) \gamma(x)- \frac{\Delta_S}{2} \sigma^z + ig \sigma^x \gamma(0) \gamma_{decoupled}.  \nonumber \tag{S32} 
\end{equation}
\noindent This is the continuum formulation of the microscopic coupling illustrated in the main text. To further elucidate the importance of the vortex degree of freedom at the injection point, we consider three possibilities. First, in the case when the dangling injection Majorana is completely decoupled, the injection vortex (corresponding to the flipped $U_{i,j}$ at the injection point) is zero energy and the procedure for TPST remains identical. Second, in the case when the injection Majorana is weakly interacting with a single nearby Majorana (respecting the interaction symmetry), the injection vortex is low-energy. In this case, the splitting of the spin-register will need to be retuned to account for the creation of this low-energy vortex and TPST will then naturally create both a fermion and an injection vortex. Crucially, tunneling of the injection vortex into the bulk will be energetically disallowed since $\Delta_v^{injection} \ll \Delta_v^{bulk}$. Problems only arise in the third case, when the injection Majorana is interacting strongly (order $\kappa$) with a single nearby Majorana, and the injection vortex is hence high-energy.  In this case, spin-edge coupling will create an injection vortex which can diffuse into the bulk; thus, upon the return of the traveling fermion, the injection vortex may no longer be localized near the injection point, causing dephasing when the quantum information is recaptured. 

\vspace{5mm}

\noindent Next, we consider the role of interactions between edge modes. The fidelity of topological state transfer will be dependent on these interactions as they induce decay of the Majorana quasiparticles. Here, we begin by estimating the lifetime of such excitations in the continuum edge setting by taking into account the leading order interaction term. We consider, at $T=0$, the situation where we tunnel a single quasiparticle excitation into the chiral edge from an associated spin register. The Hamiltonian is,
\begin{align}
H &= H_e + H'_e \nonumber \tag{S33} \\
H'_e = \lambda \int dx\, \gamma(x) &(i\partial)  \gamma(x) (i\partial)^2 \gamma(x) (i\partial)^3  \gamma(x). \nonumber
\end{align}
\noindent To evaluate the interaction induced decay rate of the quasiparticles, we use Fermi's golden rule and consider the relevant interaction matrix elements coupling an incoming excitation $\gamma_p |\Omega \rangle$ with three outgoing (decayed) excitations $\gamma_{p_1} \gamma_{p_2}\gamma_{p_3}|\Omega \rangle$. The associated decay rate takes the form,
\begin{equation}
\Gamma = 2\pi \int \frac{dp_1}{2\pi} \frac{dp_2}{2\pi} \frac{dp_3}{2\pi} | \langle \Omega | \gamma_{-p_1} \gamma_{-p_2}\gamma_{-p_3} H'_{e} \gamma_p | \Omega \rangle |^2 \delta (\epsilon_p-\epsilon_{p_1}-\epsilon_{p_2}-\epsilon_{p_3}) \nonumber \tag{S34} 
\end{equation}
\noindent where the delta-function imposes energy conservation and the integrals are one dimensional because the quasiparticle is confined to the droplet edge. To evaluate the decay rate, we begin by considering the interaction matrix element,
\begin{equation}
M = \lambda \int \frac{dk_1}{2\pi} \frac{dk_2}{2\pi} \frac{dk_3}{2\pi} \frac{dk_4}{2\pi} k_2 k_3^2 k_4^3 (2\pi) \delta (k_1 + k_2 +k_3+k_4) \langle  \gamma_{-p_1} \gamma_{-p_2}\gamma_{-p_3} \gamma_{k_1} \gamma_{k_2}\gamma_{k_3}  \gamma_{k_4} \gamma_p \rangle \nonumber \tag{S35} 
\end{equation}
\noindent where we have represented the interaction Hamiltonian in momentum space; by employing Wick's theorem, we can contract the matrix element into a function of two point Majorana correlators. The only terms from this contraction which contribute are connected terms of the form $\langle \gamma_{-p_1} \gamma_{k_1} \rangle \langle \gamma_{-p_2} \gamma_{k_2} \rangle \langle \gamma_{-p_3} \gamma_{k_3} \rangle \langle \gamma_{k_4} \gamma_{p} \rangle$ and such terms yield, $\Gamma \sim \lambda^2 p^{13}/v$. The leading order temperature correction in the limit $|vp| \gg kT$, is obtained from a Sommerfeld type expansion \cite{Sommerfeld28} and yields, 
\begin{equation}
\Gamma \sim \frac{\lambda^2 p^{13}}{v} + \frac{\lambda^2 p^{11}T^2}{v} +\mathcal{O}(T^4). \nonumber \tag{S36} 
\end{equation}
